\documentclass[12pt,a4paper]{article}
        
\usepackage{a4wide}
\usepackage{amsmath}
\usepackage{amssymb}
\usepackage{amsfonts}
\usepackage{epsfig}
\usepackage{color}
\usepackage{exscale}
\usepackage{float}
\usepackage{bbm}
\usepackage[numbers,sort&compress]{natbib}

\makeatletter
\@addtoreset{equation}{section}
\makeatother

\title{\vskip-2cm Real-Time Evolution of Strongly Coupled Fermions driven by
Dissipation}

\author{E.\ Huffman$^a$, D.\ Banerjee$^b$, S.\ Chandrasekharan$^a$, and
U.-J.\ Wiese$^c$
\\ \\
$^a$ Department of Physics, Duke University\\
Durham, North Carolina 27708, USA\\
$^b$ NIC, DESY, Platanenallee 6, D-15738 Zeuthen, Germany\\ 
$^c$ Albert Einstein Center for Fundamental Physics \\
Institute for Theoretical Physics, Bern University \\
Sidlerstrasse 5, CH-3012 Bern, Switzerland \\ \\
}

\begin{document} 

\maketitle

\vspace{-1cm}

\begin{abstract} \normalsize
We consider the real-time evolution of a strongly coupled system of lattice
fermions whose dynamics is driven entirely by dissipative Lindblad processes,
with linear or quadratic quantum jump operators. The fermion 2-point functions
obey a closed set of differential equations, which can be solved with linear
algebra methods. The staggered occupation order parameter of the $t$-$V$ model
decreases exponentially during the dissipative time evolution. The structure
factor associated with the various Fourier modes shows the slowing down of 
low-momentum modes, which is due to particle number conservation. The
processes with nearest-neighbor-dependent Lindblad operators have a decay rate that is 
proportional to the coordination number of the spatial lattice.
\end{abstract}

\newpage
 
\section{Introduction}

Understanding the real-time dynamics of large strongly coupled quantum systems
is a challenge that affects many areas of physics. In particular, simulations 
of the real-time path integral on classical computers suffer from severe 
complex action problems, which prevent the application of importance sampling 
underlying quantum Monte Carlo \cite{Tro05}. Since the unitary time evolution of large
isolated quantum systems may lead into paradoxical Schr\"odinger-cat-like 
states, it seems unlikely that their behavior can be captured by classical
computation. Quantum simulators, i.e.\ special purpose digital \cite{Llo96} or 
analog \cite{Jak98} quantum computers, are emerging as promising new tools 
that, thanks to their quantum hardware, do not suffer from sign or complex 
action problems. They can thus be used to address systems for which no 
``classical'' solution of the sign problem has been found. Quantum simulators 
have been constructed in atomic and condensed matter physics 
\cite{Cir12,Lew12,Blo12,Bla12,Asp12,Hou12}, and more recently also in a 
particle physics context \cite{Kap11,Szi11,Zoh12,Ban12,Ban13,Zoh13,
Tag13a,Tag13b,Wie13}. Since quantum simulators are not yet readily available 
for most
systems, it is interesting to ask what aspects of real-time quantum dynamics can
be captured by simulations using classical computers. First of all, a lot of
progress has been made for gapped systems with moderate entanglement in one 
spatial dimension. In this case, the matrix product states underlying 
the density matrix renormalization group (DMRG) \cite{Whi92,Sch05} provide a 
basis for simulating the real-time evolution over moderate time intervals 
\cite{Vid03,Whi04,Ver04,Zwo04,Dal04,Bar09,Piz13}. For large higher-dimensional 
quantum systems, on the other hand, no unbiased computational method exists for
addressing real-time dynamics. 

Purely dissipative dynamics driven by a Lindblad process, in which the 
Hamiltonian is discarded, plays an important role for state preparation of 
ultracold atom systems
\cite{Cor74,Ber02,Aar02,Ber08,Die08,Dal09,Die10,Mue12,Sie13,DeG13,Hor13}.
Dissipation can also serve as a resource for quantum computation 
\cite{Rau01,Nie01,Chi02,Ali04,Ver09} and for entanglement generation 
\cite{Kra11}. Recently, some severe sign problems have been
solved for strongly interacting 2-dimensional quantum spin systems that are
entirely driven by their coupling to a dissipative environment \cite{Ban14}. 
The Lindblad processes simulated so far can also be interpreted as sporadic 
randomized measurement processes, for example, of the total spin of a 
nearest-neighbor spin pair. A Lindblad dynamics with Hermitean quantum jump 
operators, which is equivalent to a measurement process, ultimately leads to a 
featureless infinite temperature density matrix. In spite of this, the 
corresponding heating process leads far out of thermal equilibrium and is 
interesting to investigate. In particular, depending on the dissipative process,
some physical quantities like the total magnetization or staggered magnetization
may be conserved. In that case, the thermalization of certain magnetization 
Fourier modes is slowed down \cite{Heb15}, and the transport of the conserved 
quantity is governed by a diffusion process. The flow of magnetization or 
staggered magnetization between a ferro- and an antiferromagnetic reservoir, 
occupying two initially separated parts of the volume, has been studied in 
\cite{Ban15}.

In this paper, we investigate Lindblad processes for fermions, with linear or
quadratic quantum jump operators. Again, we study purely dissipative dynamics
driven by the interaction with an environment, not by a Hamiltonian. Even in
Euclidean time, fermion simulations are often challenging. Due to Fermi 
statistics, negative signs enter the fermion path integral and may prevent the
application of importance sampling. In such cases, a severe fermion sign problem
arises. In specific cases, in particular at half-filling, integrating out the 
fermions may lead to a non-negative fermion determinant, such that Monte Carlo
is applicable. In other cases, for example, in the repulsive Hubbard model away
from half-filling or in lattice QCD at non-zero baryon density, a severe sign
problem prevents numerical investigations from first principles. Still, in some
interesting cases, severe fermion sign problems have been solved, for example,
using the meron-cluster algorithm \cite{Cha99} or the fermion bag approach 
\cite{Cha10, Cha12, Cha13,Huf14}. Fermion simulations in real time are particularly 
challenging, because one encounters both the fermion sign problem and the
complex weight problem characteristic for real-time processes. It is known that for some Hermitean jump operators,
analytic expressions for some correlations can be found in bosonic and fermionic systems 
and criteria for when this is possible have been developed \cite{Zun14}. 
Steady states for some of the correlations, specifically for dissipative one-dimensional 
noninteracting spin systems, have been found \cite{Pro08, Zun10, Pro10a, Pro10b}. Here we identify
specific fermionic Lindblad processes, with linear or quadratic quantum jump operators, 
for which analytic expressions for the real-time evolution of certain observables may be obtained, and we 
analyze their behavior. As an example, we use the Hamiltonian of the $t$-$V$ model to generate a 
correlated initial fermion density matrix, but the results are more general and would be applicable to any pure state 
or a more general density matrix. 

At low temperature, the $t$-$V$ Hamiltonian leads to a staggered occupation of fermions on a bipartite
lattice. The real-time process, which is driven entirely by the Lindblad 
dynamics (and not by the Hamiltonian), leads to an exponentially fast 
destruction of the staggered occupation order. In the case of quadratic quantum
jump operators, we investigate the dependence of the exponential decay rate on
the lattice geometry. As with the quantum spin systems studied in 
\cite{Ban14,Heb15,Ban15}, the thermalization of some Fourier modes of the 
fermion density is slowed down due to fermion number conservation.

The rest of the paper is organized as follows. In Section 2 we set up the
system of fermions, and in section 3 we analytically
investigate dissipative fermion dynamics with linear quantum jump operators.
Processes driven by quadratic quantum jump operators are studied in
Section 3. Finally, Section 4 contains our conclusions.

\section{The System}
We begin with the Lindblad equation, which describes the evolution of a density matrix in real-time, assuming the process is Markovian:
\begin{equation}
\partial_t \rho = -i\left[H,\rho\right] + \frac{1}{\epsilon} 
  \sum_{\{\nu,n\}} \left(L_{\nu,n} \rho L_{\nu,n}^\dagger - \frac{1}{2} L_{\nu,n}^\dagger L_{\nu,n} \rho - \frac{1}{2} \rho L_{\nu,n}^\dagger L_{\nu,n}\right).
\label{Lindbladeq}
\end{equation}
The $L_{\nu,n}$ operators are \textit{Lindblad operators}, 
or quantum jump operators, which provide an effective description of how 
the external environment acts on the quantum system under study. This equation 
gives the time-evolution equation of the quantum system with the presence of the 
Hamiltonian and the environment. This is the most general trace-preserving 
equation for the density matrix for the quantum system under study that excludes 
the back-reaction of the quantum system onto itself. There are $n$ types of Lindblad 
operators in (\ref{Lindbladeq}), and each type has $\nu$ possible degree of 
freedom labels. In our examples to come, each $\nu$ will be a spatial site 
label or a nearest neighbor pair label.

In general the Lindblad operators may be completely arbitrary, 
and thus would not be normalized, but in all of our examples they obey the relation 
\begin{equation}
\left(1-\epsilon \mu^2 \mathcal{N}\right) \mathbbm{1} + \epsilon \sum_{\nu,n} L_{\nu,n}^\dagger L_{\nu,n} =\mathbbm{1} .
\label{nor}
\end{equation}
The constant $\mu$ is the coefficient of the Lindblad operators and 
has the physical significance of indicating the strength of the coupling of the environment 
to the system per unit time. In dimensionless units, it can be expressed as $\mu^2 \tau$, 
where $\tau$ is real time. The constant $\mathcal{N}$ is the number of degree of freedom 
labels that the operators have, and $\epsilon$ is an arbitrary step in real time --- 
it is possible to take the limit $\epsilon\rightarrow 0$ smoothly, 
and define everything in in continuous real time. We normalize the Lindblad operators according 
to (\ref{nor}) because doing so simplifies the expressions we will be using below. 
For this analysis, we will consider purely dissipative systems, i.e., the Hamiltonian will
be switched off, $H=0$, for the real-time part of the evolution.

We will at times assume an initial correlated density matrix for a bipartite lattice, 
formed using the $t$-$V$ Hamiltonian,
\begin{equation}
H = \sum_{\left\langle x,y\right\rangle}\left[-t \left(c_x^\dagger c_y + {\rm h.c.}\right) + V \left(n_x - \frac{1}{2}\right) \left(n_y - \frac{1}{2}\right) \right],
\end{equation}
at low temperature. Other times we will use a completely staggered pure state as the initial state, 
consisting of particles on the even sites and holes on the odd sites.
\begin{table}
\begin{tabular}{|c|c|c|}
\hline
Name & Symbol & Expression\\
\hline
Staggered Magnetization & $O$ & $\sum_x \eta_x \left(n_x - \frac{1}{2}\right)$\\
\hline
Uniform Magnetization & $O_u$ & $\sum_x \left(n_x - \frac{1}{2}\right)$\\
\hline
Staggered Susceptibility & $\chi_s$ & $ \sum_{xy} \eta_x \eta_y \left(n_x - \frac{1}{2}\right) \left(n_y - \frac{1}{2}\right)$\\
\hline
Uniform Susceptibility & $\chi_u$ & $ \sum_{xy} \left(n_x - \frac{1}{2}\right) \left(n_y - \frac{1}{2}\right)$\\
\hline
Fourier Mode & $F\left(q\right)$ & $\sum_x \exp\left(ip_q\cdot x\right) \left(n_x - \frac{1}{2}\right)$\\
\hline
Structure Factor & $S\left(q\right)$ & $\sum_{xy} \exp\left(ip_q\cdot\left(x-y\right)\right) \left(n_x - \frac{1}{2}\right) \left(n_y - \frac{1}{2}\right)$\\
\hline
\end{tabular}
\caption{Observables referenced in this paper. $p_q = \left(2\pi q_1/L_1, 2\pi q_2/L_2, ...,2\pi q_d /L_d\right)$, where $L_1, L_2, ...,L_d$ give the number of lattice points in directions $1, 2, ...,d$ respectively.}
\label{table:tab}
\end{table}

We may measure the amount of order present in the system using the observable $O = \sum_x \eta_x \left(n_x - \frac{1}{2}\right)$, where $\eta_x$ is $+1$ for one sublattice and $-1$ for the other. Other useful observables that will be referenced in this paper are given in Table \ref{table:tab}.
To represent the real time, we use the variable $\tau$ (not to be confused with the tight-binding constant $t$). Introducing an operator $C$, for a chain of Lindblad operators,
\begin{equation}
C_{\left\{\nu,n\right\},k} = L_{\nu_k,n_k} L_{\nu_{k-1},n_{k-1}} ...  L_{\nu_1,n_1},
\end{equation}
the average observable as a function of real-time is then given by
\begin{equation}
\begin{aligned}
\left\langle O \left(\tau\right)\right\rangle &= \frac{ \mbox{Tr}\left( e^{- \mu^2 \mathcal{N} \tau} \sum_{\left\{\nu, n \right\},k} \int_0^{\tau_2} ...\int_0^{\tau} d\tau_1...d\tau_k C_{\left\{\nu,n\right\},k}^\dagger O C_{\left\{\nu,n\right\},k} e^{-\beta H}\right)}{\mbox{Tr}\left( e^{-\mu^2 \mathcal{N} \tau} \sum_{\left\{\nu, n \right\},k} \int_0^{\tau_2} d\tau_1...\int_0^\tau d\tau_k C_{\left\{\nu,n\right\},k}^\dagger C_{\left\{\nu,n\right\},k} e^{-\beta H}\right)} \\
&=\qquad\qquad\qquad\qquad\quad\frac{ \mbox{Tr}\left( O\left(\tau\right) e^{-\beta H}\right)}{\mbox{Tr}\left( e^{-\beta H}\right)}.
\end{aligned}
\label{obs}
\end{equation}
Here the sum over $\left\{\nu, n\right\}$ indicates a summation over all possible $\nu_1, \nu_2,...,\nu_k$ combinations,
as well as a sum over all possible $n_1,n_2,...,n_k$ combinations for (all possible) $k$ insertions. The last line in (\ref{obs}) defines $O\left(\tau\right)$. The simplified denominator may be obtained using the identity $\sum_{\{\nu,n\}} L_{\nu,n}^\dagger L_{\nu,n} = \mu^2 \mathcal{N} \mathbbm{1}$. We see that indeed the real-time part of the denominator will simplify in the following way:
\begin{equation}
e^{-\mathcal{N} \mu^2 \tau} \left(\mathbbm{1} + \tau \mathcal{N} \mu^2 \mathbbm{1} + \frac{\tau^2}{2} \mathcal{N}^2 \mu^4 \mathbbm{1} + ...\right) =  \mathbbm{1}.
\label{norm}
\end{equation}
Here we note that (\ref{obs}) holds only if the normalization condition (\ref{nor}) for the Lindblad operators is met. Otherwise we would need one additional type of insertion, the Krauss operator $M_0 = 1 - \epsilon \sum_{\nu,n} L^\dagger_{\nu,n} L_{\nu,n}$ to ensure full normalization. We note in passing that our integral approach is more suited to simulations. Analytical results may also be obtained by solving the equivalent differential equation for the real-time evolution for the operators in real-time, given by $d_\tau {\rm Tr}\left[O \rho\left(\tau\right)\right] = \frac{1}{2} \sum_{\nu,n} {\rm Tr}\left[\left(L^\dagger_{\nu,n} \left[O, L_{\nu,n}\right] + \left[L^\dagger_{\nu,n}, O\right] L_{\nu,n}\right) \rho\right]$, which holds for any set of Lindblad operators, normalized or not.

We will now consider several sets of Lindblad operators for which the numerator may also be summed analytically.
\section{Dissipative Fermion Dynamics with Linear Quantum Jump Operators}
The first set of linear quantum jump operators will be limited to one site for each operator, and the second set involves nearest neighbor pairs on the lattice. We also note that according to \cite{Zun14} it is possible to find analytic solutions for linear fermionic Lindblad operators. This knowledge, however does not tell us what to expect in particular for the steady state behavior at large times for each linear Lindblad system.
\subsection{Single Site Linear Quantum Jump Operators}
We consider the following two types of operators:
\begin{equation}
\begin{array}{cc}
L_{x,+}  =  \mu c_x^\dagger , \quad & L_{x,-}  =  \mu c_x .
\end{array}.
\end{equation}
Here $x$ is the lattice site index. The operators obey the relation $\left(1- \epsilon \mu^2 N\right) \mathbbm{1} + \epsilon \sum_{\{x,s\}} L_{x,s}^\dagger L_{x,s} = \mathbbm{1}$, where $N$ is the number of sites and $s\in\left\{+,-\right\}$. Here $N$ plays the part of $\mathcal{N}$ in equations (\ref{obs}) and (\ref{norm}). The constant $\mu^2$ is the probability per unit time for each jump that these operators initiate.

The surprising result is that for some observables the real-time portion of the numerator may also be summed analytically. We let $O_y = n_y - \frac{1}{2}$ and consider
\begin{equation}
\sum_{s} L_{x,s}^\dagger O_y L_{x,s}.
\end{equation}
When $x\neq y$, $O_y$ commutes with the Lindblad operators and the result is clearly $\mu^2 O_y$. For $x=y$, we have
\begin{equation}
\begin{aligned}
\sum_{s} L_{x,s}^\dagger O_y L_{x,s} & =  \mu^2 c_y^\dagger \left(n_y - \frac{1}{2} \right) c_y + \mu^2 c_y \left(n_y - \frac{1}{2} \right) c_y^\dagger \\
& =  \mu^2 \left(\frac{1}{2} - n_y \right) c_y^\dagger c_y + \mu^2 \left(\frac{1}{2} - n_y \right) c_y c_y^\dagger \\
 & =  \mu^2 \left(\frac{1}{2} - n_y\right) = -\mu^2 O_y.
 \end{aligned}
\end{equation}
From here we can see that
\begin{equation}
\sum_{\{x,s\}} L_{x,s}^\dagger O_y L_{x,s} = \mu^2 \left(N-1\right) O_y - \mu^2 O_y  = \left(N-2\right) \mu^2 O_y,
\end{equation}
and thus the sum in the numerator of (\ref{obs}) is
\begin{equation}
\begin{aligned}
e^{-N \mu^2 \tau} \sum_{\left\{x ,s\right\},k} \int_0^{\tau_2} ...&\int_0^{\tau} d\tau_1...d\tau_k L_{x_1,s_1}^\dagger L_{x_2,s_2}^\dagger ... L_{x_k,s_k}^\dagger O L_{x_k,s_k} ... L_{x_2,s_2} L_{x_1,s_1} \\
&= e^{-N \mu^2 \tau} \left(\mathbbm{1} + \tau \mu^2 \left(N - 2\right) \mathbbm{1} + \frac{\tau^2}{2} \mu^4 \left(N-2\right)^2 \mathbbm{1} +...\right)  O\\
&= e^{- N \mu^2 \tau} e^{\left(N-2\right) \mu^2 \tau} O = e^{ -2\mu^2 \tau} O.
\end{aligned}
\end{equation}
Therefore, we see that the following holds:
\begin{equation}
\left\langle O\left(\tau\right)\right\rangle = e^{-2\mu^2 \tau} \frac{Tr\left(O e^{-\beta H}\right)}{Tr\left(e^{-\beta H}\right)}.
\end{equation}
We conclude that the speed of the decay for the order parameter in the system is exponential in time. The lattice size and geometry are irrelevant to the speed of the decay, with only the $\mu^2$ constant determining it. Using the same techniques, we can find an analytic value for the real-time evolution of another average observable, the staggered susceptibility $\chi_s$, defined in Table \ref{table:tab}.
In real-time, we can write $\left\langle \chi_s\left(\tau\right)\right\rangle = \mbox{Tr}\left(\chi_s\left(\tau\right)e^{-\beta H}\right)/\mbox{Tr}\left(e^{-\beta H}\right)$, where
\begin{equation}
\chi_s\left(\tau\right) = e^{-4\mu^2 \tau} \sum_{x\neq y} \eta_{x} \eta_{y} O_x O_y  + \frac{N}{4}.
\label{linchis1}
\end{equation}
From here we can see that the dominant behavior is a decay at twice the rate of that for the staggered magnetization, as expected.

Finally, we can also get an analytic expression for the evolution of the Fourier modes, $F\left(q\right)$ and structure factors $S\left(q\right)$.
From the previous proof, it follows directly that
\begin{equation}
\begin{array}{cc}
\begin{aligned}
F\left(q,\tau\right) &= e^{-2\mu^2 \tau} F\left(q\right)\\
S\left(q,\tau\right)  &= e^{-4\mu^2 \tau} \sum_{x\neq y}  \exp\left(i p_q \cdot\left(x-y\right)\right) O_x O_y + \frac{N}{4},
\end{aligned}
\end{array}
\end{equation}
and so the decay rates are independent of mode.
\subsection{Nearest Neighbor Linear Quantum Jump Operators}
The second set of linear Lindblad operators consists of,
\begin{equation}
\begin{array}{cc}
L_{xy,1}  =  \frac{\mu}{2} \left(c_x + c_y^\dagger\right) , \quad & L_{xy,2}  =  \frac{\mu}{2} \left(c_x - c_y^\dagger\right) .
\end{array}
\end{equation}
Here $x$ and $y$ are nearest neighbors, the index $n=1,2$, and $L_{yx,n}$ is counted as a separate operator from $L_{xy,n}$. There are $N$ sites total in the lattice, and we use $m$ to denote the number of nearest neighbors. The Lindblad condition holds:
\begin{equation}
\begin{aligned}
\sum_{\left\langle x,y \right\rangle,n} \left[L_{xy,n}^\dagger L_{xy,n} + L^\dagger_{yx,n} L_{yx,n} \right] = \mu^2 \frac{Nm}{2}\cdot \mathbbm{1} .
 \end{aligned}
\end{equation}
There are $N\cdot m/2$ nearest neighbor pairs, so the $\mathcal{N}$ constant in equation (\ref{obs}) is $N\cdot m / 2$. Using $c_x O_x = -O_x c_x$ and $c_x^\dagger O_x = - O_x c_x^\dagger$, we find 
\begin{equation}
\sum_n L_{xy,n}^\dagger \left(n_x - \frac{1}{2}\right) L_{xy,n} + L^\dagger_{yx,n} \left(n_x - \frac{1}{2}\right) L_{yx,n} = 0,
\end{equation}
where again the $n$ variable runs over $1,2$. With this result, we may now calculate the more comprehensive sum,
\begin{equation}
\sum_{\left\langle x,y \right\rangle, n}\left[ L_{xy,n}^\dagger O_k L_{xy,n}+ L_{yx,n}^\dagger O_k L_{yx,n}\right] = \mu^2 \left(\frac{m N}{2} - m\right) O_k .
\end{equation}
And so there is an analytic expression for the real-time evolution of $O$, given by
\begin{equation}
O\left(\tau\right) = e^{- m\mu^2  \tau} O .
\end{equation}
The rate of decay is proportional to the number of nearest neighbors per site in the lattice geometry.

Similarly, we can show that the staggered susceptibility goes as
\begin{equation}
\chi_s\left(\tau\right) = e^{-2 m\mu^2  \tau} \sum_{ x\neq y } \eta_x \eta_y O_x O_y + \frac{N}{4}.
\label{linchis2}
\end{equation}
Finally, the more general Fourier modes and structure factors for the square lattice evolve in a straightforward generalized way of how the staggered magnetization and staggered susceptibility evolve, as
\begin{equation}
F\left(q,\tau\right) = e^{-m\mu^2  \tau} F\left(q\right),
\end{equation}
and
\begin{equation}
S\left(q,\tau\right) = e^{-2 m\mu^2  \tau} \sum_{ x\neq y} \exp\left(ip_q\cdot\left(x-y\right)\right) O_x O_y  + \frac{N}{4}.
\end{equation}
As in the last set of linear Lindblad operators, there is no special $q$-dependence in the decay rate of these Fourier modes.

\section{Dissipative Fermion Dynamics with Quadratic Quantum Jump Operators}
We now turn to a set of quadratic Lindblad operators,
\begin{equation}
\begin{array}{ccc}
L_{xy,1}  =  \mu c_x^\dagger c_y , \quad & L_{xy,2}  =  \mu c_y^\dagger c_x, \quad & L_{xy,3} = \mu\left(1-n_x - n_y\right).
\end{array}
\end{equation}
Here $x$ and $y$ are nearest neighbors, and this time $L_{xy,n}$ is counted as the same operator as $L_{yx,n}$. Again the lattice has $N$ sites and there are $m$ nearest neighbors per site. There are thus $m\cdot N/2$ unique site labels total. These operators satisfy the condition:
\begin{equation}
L_{xy,1}^\dagger L_{xy,1} + L_{xy,2}^\dagger L_{xy,2} + L_{xy,3}^\dagger L_{xy,3} = \mu^2 \mathbbm{1}.
\end{equation}
For this set of operators, there is the following interesting result:
\begin{equation}
\sum_n L^\dagger_{xy,n} O_x L_{xy,n} = \mu^2 \eta_x O_y.
\end{equation}
From here, we may sum over all $\left\langle x,y\right\rangle$ to find that
\begin{equation}
\begin{aligned}
\sum_{n,\left\langle x,y \right\rangle} L_{xy,n}^\dagger & O L_{xy,n} \\
& = \sum_{n,\left\langle x,y\right\rangle,k} L^\dagger_{xy,n} \eta_k O_k L_{xy,n} \\
& = \sum_{k} \mu^2 \left(\frac{N \cdot m}{2} - 2m\right) \eta_k O_k .
 \end{aligned}
\end{equation}
From here again it is possible to extract the real-time evolution of $O$, as
\begin{equation}
 O\left(\tau\right)= e^{-2m\mu^2 \tau} O,
\end{equation}
and the rate of decay is proportional to the number of nearest neighbors in the lattice. 

This result is similar to that for the nearest neighbor linear operators, but we encounter more complexity in the Fourier modes, the susceptibility, and the structure factors. The Fourier modes still are described by a simple expression and are given (for $q_1 = q_2$) by
\begin{equation}
F\left(q,\tau\right) = e^{-m\mu^2\left(1-\cos p_q a\right)\tau} F\left(q\right).
\end{equation}
On the other hand, finding the staggered susceptibility is more complicated. In the previous examples of linear operators, $O_x O_y$ would remain proportional to itself once we summed over all the sites for the three possible Lindblad insertions, but for the quadratic operators, there is mixing between $O_x O_y$ and $O_x O_{y+\alpha}$, where $\alpha$ connects a site to one of its nearest neighbors.

The way forward is to find a way to write the susceptibility in terms of operators that remain proportional to themselves when the three possible Lindblad insertions are summed over. We define a coefficient vector $\vec{v}$ with entries $v_{xy}$, where each $v_{xy}$ represents the coefficient of a particular $O_x O_y$ term. We could start then with a vector consisting of all ones, and this would represent the uniform susceptibility observable $\chi_u$. We then construct a matrix $M$ such that ${\vec{v}\,}'=M\vec{v}$ is a new vector resulting from the action of $\sum_{\left\langle x,y\right\rangle ,n} L^\dagger_{xy,n} \chi_u L_{xy,n}$. Each entry of ${\vec{v}\,}'$ now has a new coefficient that comes from the action of the Lindblad operators.

The matrix $M$ is symmetric, and so may be diagonalized. The eigenvectors of this matrix are useful because they can be interpreted as lists of coefficients for new $E_{\lambda_i}$ ``eigen-operators.'' We find that $\sum_{\left\langle xy\right\rangle,n} L^\dagger_{xy,n} E_{\lambda_i} L_{xy,n} = \lambda_i E_{\lambda_i}$, where $\lambda_i$ is the corresponding eigenvalue for the particular eigenvector, and thus for the operator $E_{\lambda_i}$, we can easily find the time evolution,
\begin{equation}
E_{\lambda_i} \left(\tau\right) = e^{-\mu^2\left(\frac{Nm}{2}-\lambda_i\right)\tau} E_{\lambda_i} .
\end{equation}
We see now that if we can write the $\chi_s$ operator in terms of these conserved operators $E_{\lambda_i}$, we obtain an exact expression for the time evolution of the system. A special case is the uniform susceptibility, $\chi_u$, which commutes with the Lindblad insertions, and thus will always be a conserved operator. Because of the commutation, its corresponding coefficient eigenvector will always have an eigenvalue of $Nm/2$, resulting in no decay for the observable in time.

\begin{figure}
\begin{center}
\includegraphics{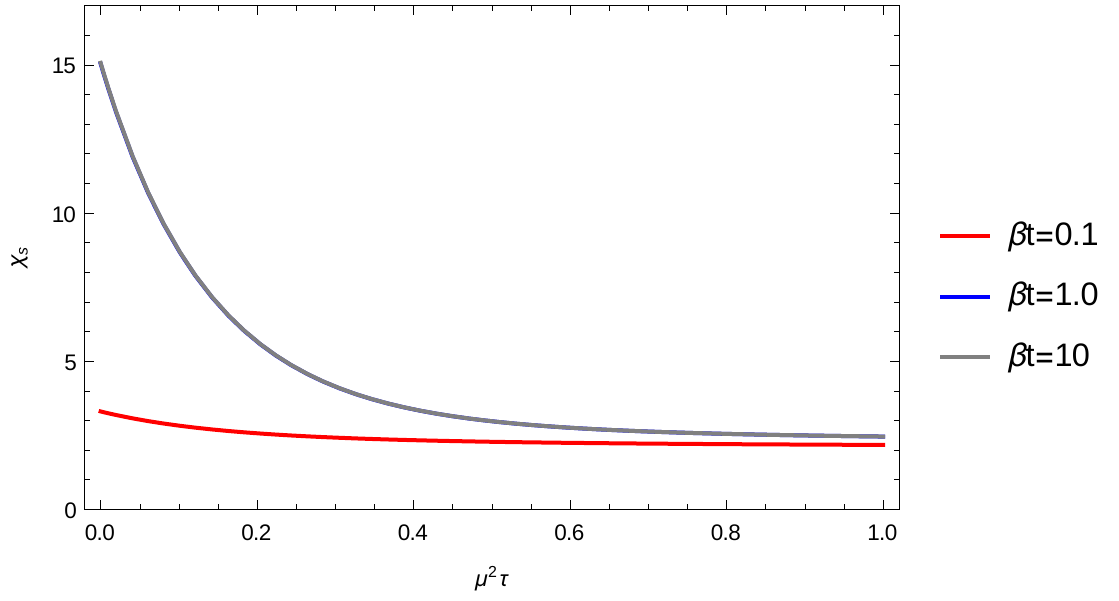}
\caption{The real-time evolution of the staggered susceptibility $\left\langle \chi_s\right\rangle$ for a one-dimensional eight site system. The initial state is prepared using the $t$-$V$ model with parameters $t=1.0$ and $V=10.0$. The $\beta$ values are $1/T$, where $T$ is the temperature of the system. We set $\mu^2 = 1.0$ and use $\tau$ to represent the real time. The line for $\beta t = 10$ is almost directly on top of the line for $\beta t = 1.0$. This is why only two curves are visible.}
\label{imag}
\end{center}
\end{figure}

Because $M$ is symmetric, we will get a set of eigenvectors that fully span their space, and so we can always form $\chi_s$ out of some linear combination of the operators $E_{\lambda_i}$. If we represent the coefficients of this linear combination of operators as $\psi_1,\psi_2,...,\psi_{N^2}$, we have
\begin{equation}
\chi_s = \psi_1 E_{\lambda_1} + \psi_2 E_{\lambda_2} + ... + \psi_{N^2} E_{\lambda_{N^2}}.
\end{equation}
The real-time evolution of the staggered susceptibility operator is then
\begin{equation}
\begin{aligned}
\chi_s\left(\tau\right) = \psi_1 & e^{-\mu^2\left(\frac{Nm}{2}-\lambda_1\right)\tau} E_{\lambda_1} + \psi_2 e^{-\mu^2\left(\frac{Nm}{2}-\lambda_2\right)\tau} E_{\lambda_2}  \\
&+ ... + \psi_{N^2} e^{-\mu^2\left(\frac{Nm}{2}-\lambda_{N^2}\right)\tau} E_{\lambda_{N^2}}.
\end{aligned}
\end{equation}
Thus we have proven that the behavior of the staggered susceptibility in time will always be described exactly as a sum of exponentially decaying functions. Figure \ref{imag} shows the real-time evolution of the quantity $\left\langle \chi_s \right\rangle$ in time, for a linear system whose initial state was prepared using the $t$-$V$ model. For high values of $\beta$, we see high initial values of the staggered susceptibility, showing that the initial state is highly ordered.

\begin{figure}[ht]      
  \begin{center} 
    \includegraphics[width=7cm]{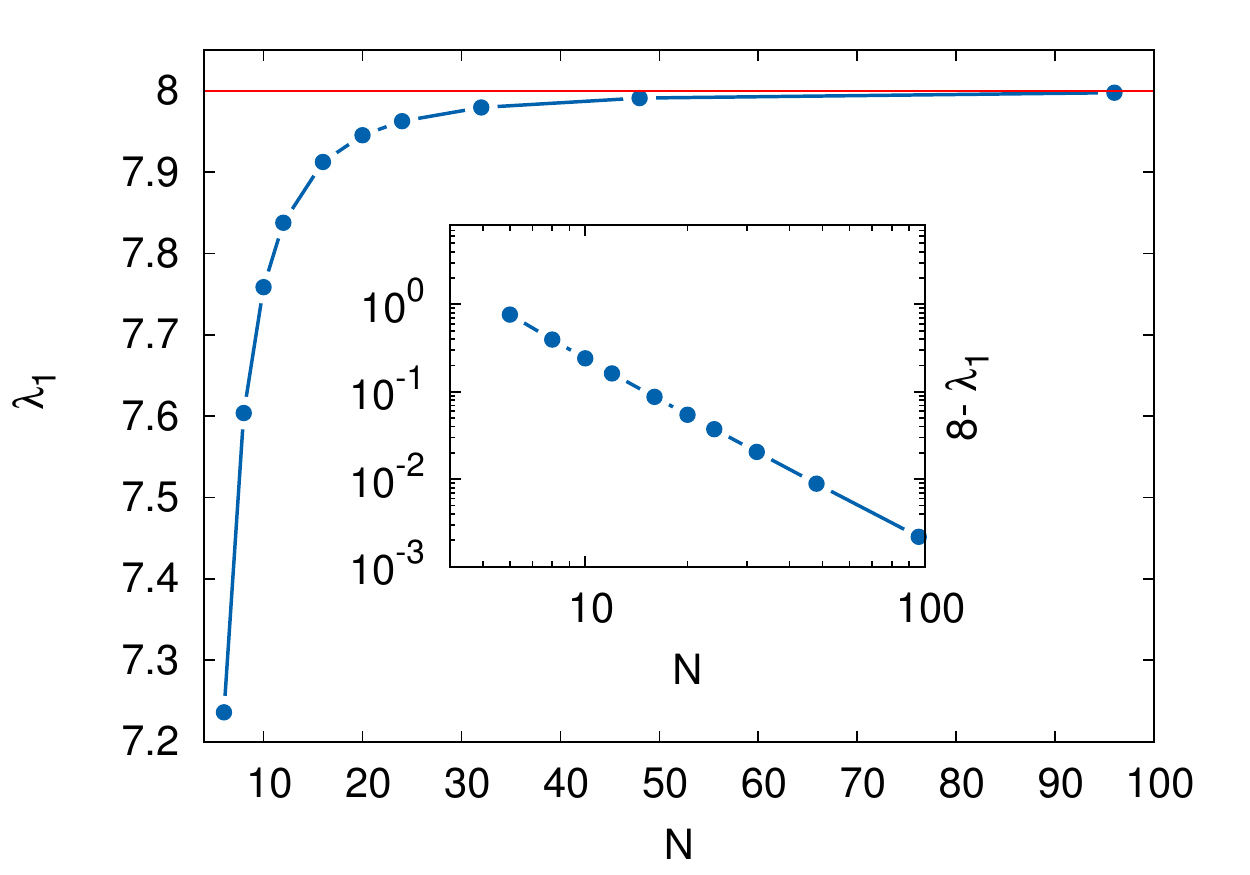}
    \includegraphics[width=7cm]{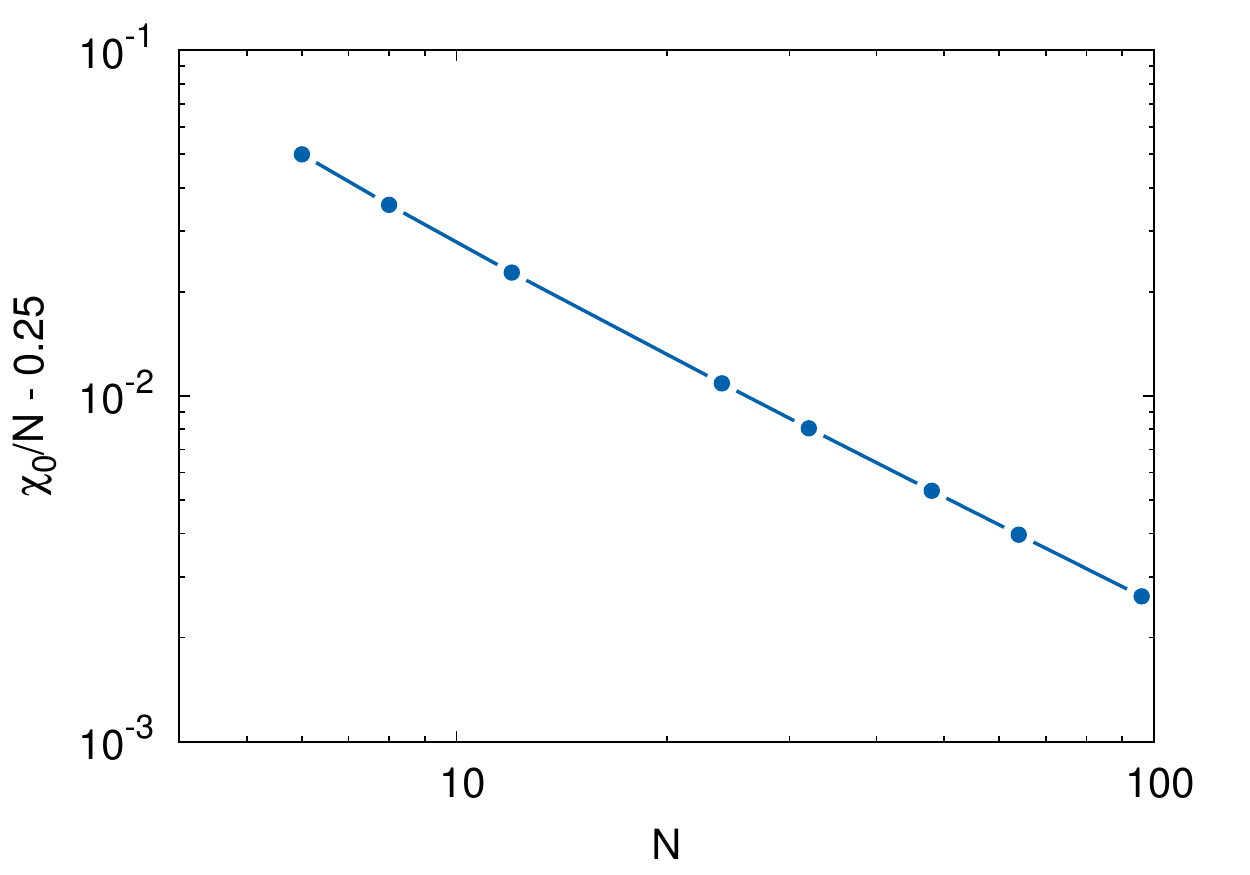}
   \caption{In these plots we begin with a completely staggered state in a one-dimensional lattice with $N$ sites. The left figure shows how the dominant operator approaches a real-time decay rate of $4\mu^2 m$ as lattice size increases. We set $\mu=1.0$ and the lattices are linear, so $m=2$ and thus the decay rate approaches $8.0$. The right figure shows how $\chi_0/N$ approaches $1/4$ as a function of $N$.}
   \label{compare}
  \end{center}
\end{figure}

While the evolution for this system is more complicated than for previous systems, we do find for each lattice size there is one particular operator component (which we will set as $E_{\lambda_1}$) which obeys a staggered ordering and makes a larger contribution to $\chi_s$ than any of the other components. The dominant operator's contribution is made more dominant the more highly ordered the system is, and we find that as the lattice size increases, the eigenvalue $\lambda_1$ for this dominant operator approaches $\frac{Nm}{2} - 4m$, resulting in a decay rate for the dominant observable approaching $-4m\mu^2$. This is consistent with what we would expect because the susceptibility is the square of the magnetization, which had a decay rate of $-2m\mu^2$. The plot to the left in figure \ref{compare} shows this approach as the lattice size increases.

As we see from equations (\ref{linchis1}) and (\ref{linchis2}), for both sets of linear Lindblad operators the staggered susceptibility decays to the constant $N/4$. However, for the quadratic Lindblad operators the constant term reaches N/4 only in the large N limit. In the right plot of figure \ref{compare}, we show the $N$-dependence of this constant term for quadratic Lindblad operators. Interestingly, these finite-volume effects occur in the quadratic case but not in the linear cases.

\begin{figure}
\begin{center}
\includegraphics{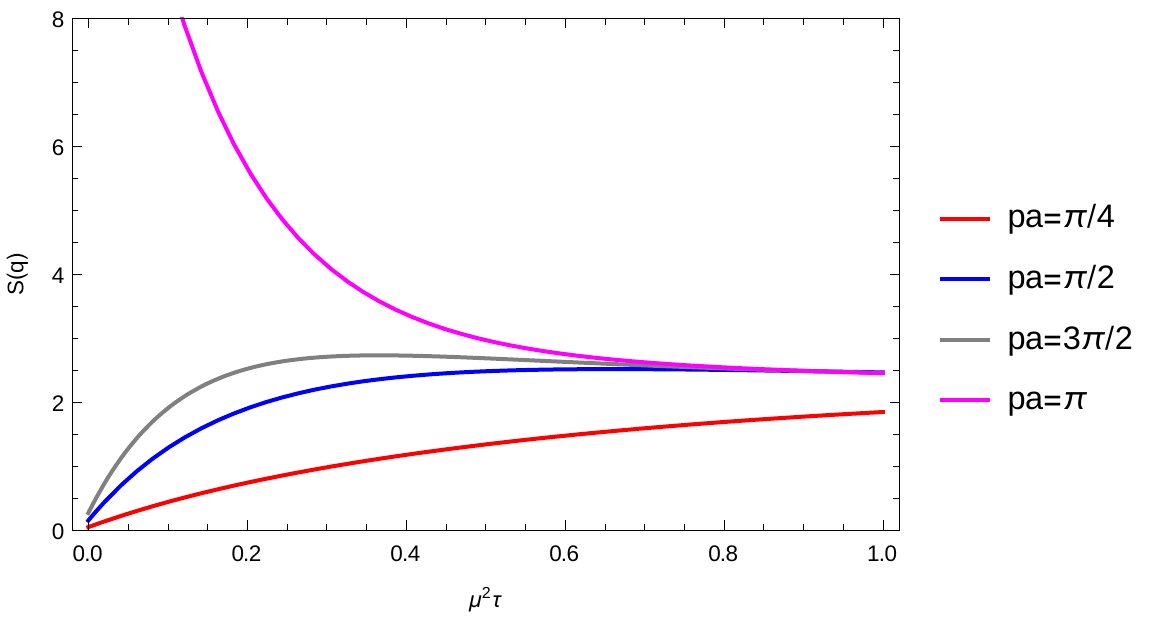}
\caption{The evolution of the structure factors for a state prepared using the $t$-$V$ model with $t=1.0$, $V=10.0$, and $\beta = 10.0$. The lattice is one-dimensional and has eight sites.}
\label{structure}
\end{center}
\end{figure}

We can next use these same operators $E_{\lambda_i}$ to form the structure factors and find their time evolution. Figure \ref{structure} shows the real-time evolution of these structure factors. We now clearly see a difference in time-dependence for different values of $q$. Low-momentum Fourier modes are slowed down, due to the fact that the zero-mode, i.e. the total particle number, is conserved.
\begin{figure}[ht]       
    \includegraphics[width=7cm]{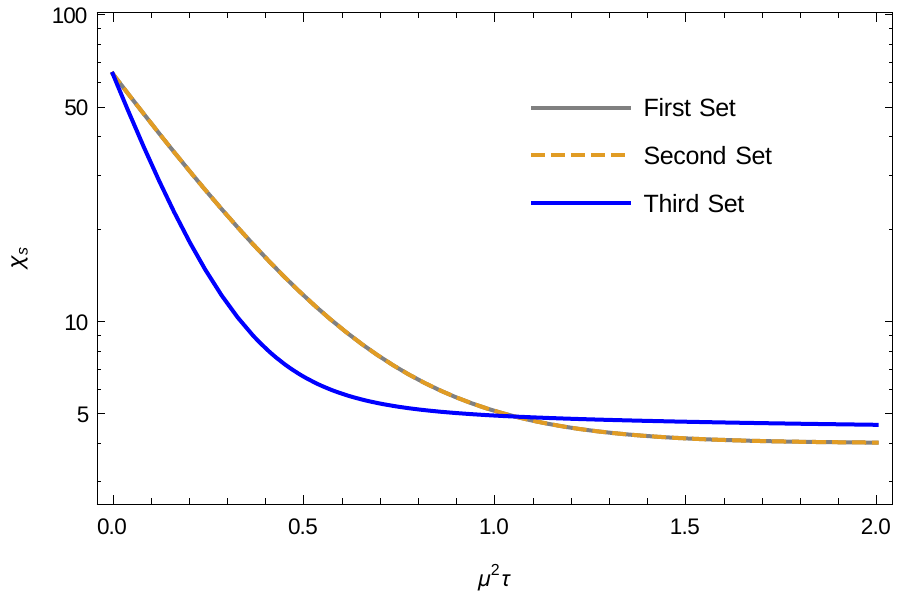}   
    \hspace{10px}
    \includegraphics[width=7cm]{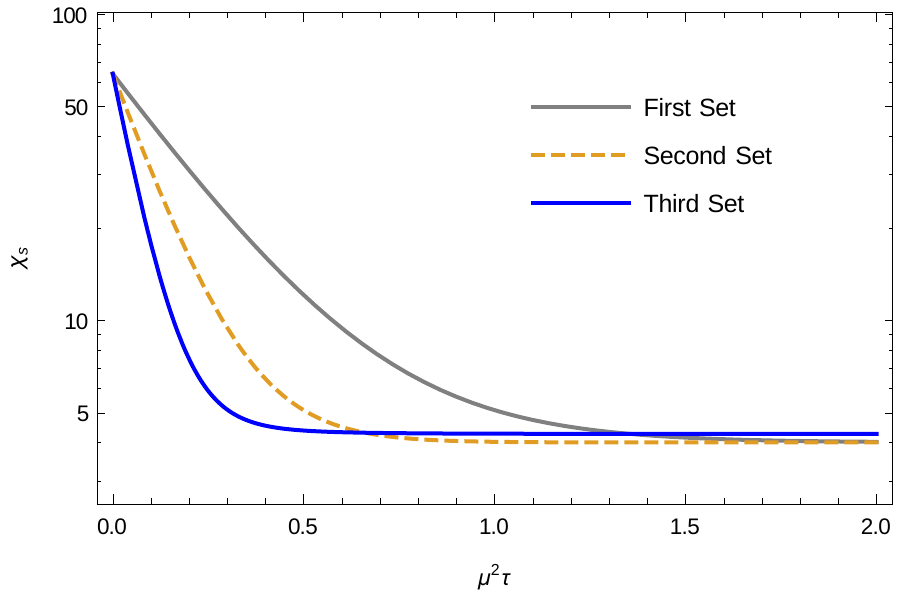}
    \caption{We begin with a pure completely staggered state. To the left we see the real-time decays of the staggered susceptibilities for a sixteen site linear lattice, and to the right we see the real-time decays of the staggered susceptibilities for a $4 \times 4$ square lattice.}
        \label{approach}
\end{figure}

Finally, we compare the decay rates of the susceptibilities for all three types of Lindblad operators in Figure \ref{approach}. We use two different lattice geometries: a $16$ site linear lattice and a $4 \times 4$ square lattice, to capture the full comparison of their time evolution. In the linear case, the $\left\langle \chi_s\right\rangle$ for the first two linear Lindblad operator sets exhibit the same decay behavior, with the quadratic operators causing a faster decay and a different limiting value $\chi_0$. In the $4 \times 4$ square lattice, we then see the difference between the first two Lindblad operators manifest itself, since the second Lindblad set has a decay rate that grows with the number of nearest neighbors, while the first Lindblad set has a constant decay rate. As for the third set, it again has the fastest decaying $\left\langle \chi_s \right\rangle$ and interestingly has the same limiting $\chi_0$ value whether the sixteen lattice sites are arranged in a line or a square.

\section{Conclusions}
In this work we have shown that it is possible to find the real-time evolution of the staggered order parameter analytically for several fermionic systems, both for linear and quadratic jump operators. We note that there are other processes higher than quadratic for which this method will also work. For example,
\begin{equation}
\begin{array}{cc}
L_{xy,1}  =  \mu c_x n_x n_y , \quad & L_{xy,2}  =  \mu c_x^\dagger n_x n_y \quad \\ L_{xy,3}  =  \mu n_x \left(1-n_y\right) \quad & L_{xy,4}  =  \mu \left(1-n_x\right) n_y
\end{array},
\end{equation}
which has a magnetization which remains constant in time, but a susceptibility that will contain some decaying modes.

Another process that can be studied is
\begin{equation}
\begin{array}{cc}
L_{xy,1}  =  \mu c_x n_x n_y , \quad & L_{xy,2}  =  \mu c_x^\dagger n_x n_y \quad \\ L_{xy,3}  =  \mu c_y^\dagger c_x n_x \left(1-n_y\right) \quad & L_{xy,4}  =  \mu c_x^\dagger c_y \left(1-n_x\right) n_y
\end{array},
\end{equation}
which will have a decaying magnetization with decay rate $2m\mu^2$, and also has a simple susceptibility decay rate of $4m\mu^2$.

All the processes involving nearest neighbor pairs caused dynamics that were dependent on the lattice geometry (specifically number of nearest neighbors), while the process involving individual sites was independent of lattice geometry. Additionally, the considered Lindblad processes all involved the same number of creation and annihilation operators acting with the same strength, though this is not necessarily a requirement. Finally, the key characteristic of our order parameter $O_i$ was that
\begin{equation}
\begin{array}{cc}
c_x O_x = - O_x c_x  \quad & c_x^\dagger O_x = - O_x c_x^\dagger.
\end{array}
\label{cond}
\end{equation}
For any other operator which satisfies (\ref{cond}), the results of this paper also hold. We stress that these results are applicable to describe the time evolution of the order parameter regardless of the initial state of the density matrix.

\section*{Acknowledgments}

The research leading to these results has received funding 
from the Schweizerischer Na\-tio\-nal\-fonds and from the European Research 
Council under the European Union's Seventh Framework Programme 
(FP7/2007-2013)/ ERC grant agreement 339220 and U.S. Department of Energy, Office of Science, Nuclear Physics program under Award Number DE-FG02-05ER41368. D. Banerjee wishes to
acknowledge interesting discussions with Arnab Sen.

\bibliographystyle{plain}

\end{document}